# Dynamic Simultaneous Multithreaded Architecture

Daniel Ortiz-Arroyo and Ben Lee

Department of Electrical and
Computer Engineering
Oregon State University
{dortiz, benl}@ece.orst.edu

**Abstract**

This paper presents the *Dynamic Simultaneous Multithreaded Architecture* (DSMT). DSMT efficiently executes multiple threads from a single program on a SMT processor core. To accomplish this, threads are generated dynamically from a predictable flow of control and then executed speculatively. Data obtained during the single context non-speculative execution phase of DSMT is used as a hint to speculate the posterior behavior of multiple threads. DSMT employs simple mechanisms based on state bits that keep track of inter-thread dependencies in registers and memory, synchronize thread execution, and control recovery from misspeculation. Moreover, DSMT utilizes a novel greedy policy for choosing those sections of code which provide the highest performance based on their past execution history. The DSMT architecture was simulated with a new cycle-accurate, execution-driven simulator. Our simulation results show that DSMT has very good potential to improve SMT performance, even when only a single program is available. However, we found that dynamic thread behavior together with frequent misspeculation may also produce diminishing returns in performance. Therefore, the challenge is to maximize the amount of thread-level parallelism that DSMT is capable of exploiting and at the same time reduce the frequency of misspeculations.

**Keywords**: Multithreading, SMT, superscalar processors, speculative execution, thread level parallelism. 1. Introduction

## 1. Introduction

Modern superscalar processors exploit instruction-level parallelism (ILP) by executing multiple instructions per cycle from a single program. In order to maximize the exploitation of ILP within a program, superscalar processors expose *true dependencies* by register renaming and attempt to mitigate the effects of *control dependencies* by performing speculative execution. Unfortunately, control-flow and true dependencies limit the amount of ILP that superscalar processors are able to exploit. To overcome this limitation, a significant amount of research has been directed at exploiting *Thread Level Parallelism* (TLP) [8, 11, 13, 14, 20, 22]. TLP can compensate for a lack of per thread ILP, and thus, both types of parallelisms can be exploited to improve the performance.

In order to exploit TLP, threads can be extracted from a single program or multiple programs. For the latter case, each program represents a thread and the threads are executed simultaneously to fully utilize a processor's resources. This technique, called Simultaneous Multithreading (SMT) [25], converts TLP into ILP to accommodate variations in ILP within a program. However, despite many advantages of SMT [5], it does nothing to improve the performance of a single program. Moreover, SMT may cause cache pollution because multiple different threads are competing for the shared cache [4].

On the other hand, extracting threads from a single program can be done either statically or dynamically. Static methods employ techniques such as parallelizing compilers or binary annotators [6] to identify threads. However, static methods are unable to provide binary compatibility. For example, parallelizing compilers become useless if the original source code is unavailable, and binary annotators tend to exploit fine-grain parallelism (e.g., basic blocks and inner loops) but in some cases require modifications to the instruction set architecture (ISA). In contrast, dynamic methods employ speculative techniques to detect and extract threads at run-time. In these methods, threads are created from loops, subroutines, and/or exception handling routines [1, 2, 11, 9, 23]. However, inter-thread dependencies occur through registers and memory when threads are not completely independent (i.e., parallel), and thus must be properly resolved to preserve the program semantics.

Based on the aforementioned discussion, this paper proposes a new architecture called *Dynamic Simultaneous Multithreading* (DSMT). DSMT is capable of functioning as a SMT processor when multiple independent programs are executed. Moreover importantly, DSMT overcomes one of the main drawbacks of SMT by efficiently exploiting both TLP and ILP from a single program. To accomplish this, DSMT employs speculative mechanisms to dynamically generate threads from the predictable behavior of loop iterations. Register values are predicted, and inter-thread dependencies are detected and resolved at run-time. Data obtained during the non-speculative execution phase of DSMT is used as a hint to speculate the posterior behavior of multiple threads. In contrast to other similar architectures, DSMT employs a simple mechanism based on state bits to keep track of inter-thread dependencies in registers and memory, synchronize thread execution, and to recover from misspeculation. Moreover, DSMT utilizes a novel greedy policy to choose those sections of code that provide the highest performance based on their past execution history. To assess the performance of DSMT, a new cycle-accurate, execution-driven simulator called *DSMTSim* was developed. This simulator is capable of reproducing in detail the complex dynamic behavior of DSMT.

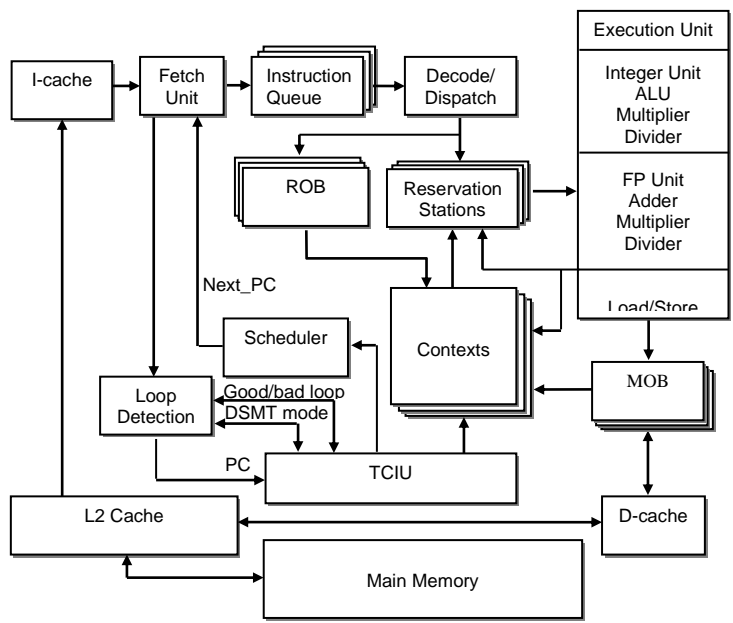

Figure 1: DSMT's Microarchitecture.

This paper is organized as follows. Section 2 describes architectures closely related to DSMT. Section 3 provides a detailed description of the DSMT microarchitecture, including details of how threads are generated and spawned and the mechanisms used to keep track of inter-thread dependencies in registers and memory. The discussion of DSMTsim and the simulation results are presented in Section 4. Finally, Section 5 concludes the paper and discusses future expansions to DSMT.

## 2. Related Work

The proposed architecture for DSMT was drawn from a plethora of related works that studied different ways of exploiting both ILP and TLP from a single program [1, 2, 6, 11, 14 20, 23, 29]. These proposals share many similarities on how thread-level speculation is supported and basically differ on how much of this support is provided in hardware versus software. Since static techniques for exploiting TLP using binary annotators or parallelizing compilers have already been discussed in [6, 22, 19], this section briefly describes architectures similar to DSMT that dynamically exploit TLP.

In Clustered Multithreaded Architecture (CMA), a control speculation mechanism dynamically identifies threads from different iterations of a loop. These threads are then executed concurrently on several thread units [9, 11]. Thread units are interconnected through a ring topology and iterations are allocated to thread units based on their execution order. Each thread unit has its own physical register, register map table, instruction queue, functional units, local memory, and reorder buffer. Inter-thread data dependencies through registers and memory are predicted with the help of a history table called the *loop iteration* table. When a speculative thread is created, its logical register file and its register map table are copied from its predecessor. At the same time, the *increment predictor* will initialize any live and predictable register. When a thread finishes, its output predictions are verified and mispredictions are handled by selective re-execution. Inter-thread memory dependence speculation is performed by means of a *multi-value* cache. This cache memory stores, for each address, as many different data words as the number of thread units.

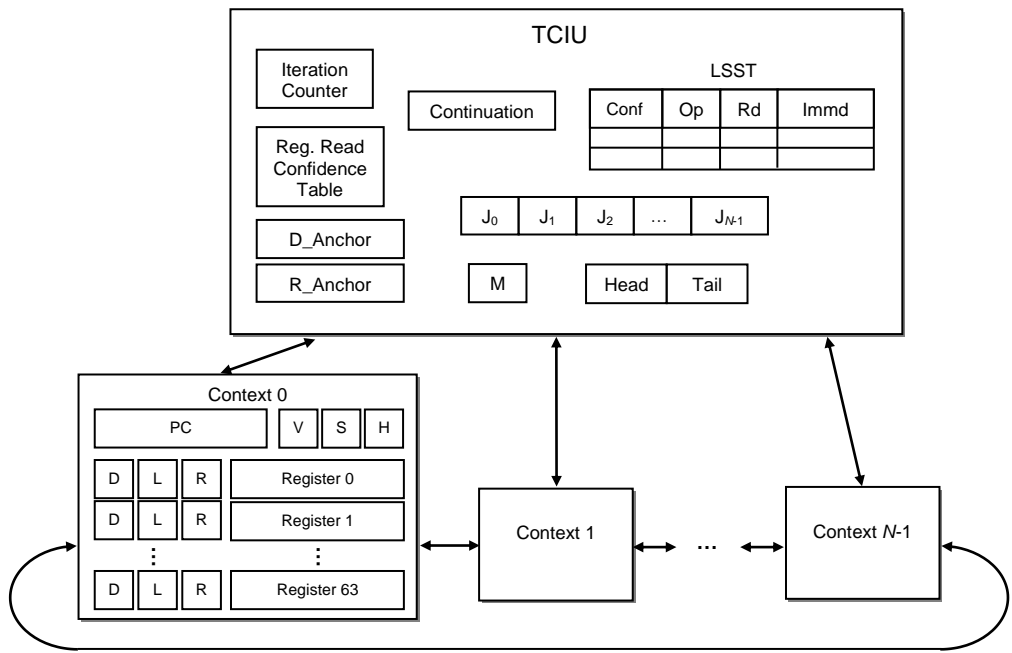

Figure 2: TCIU and Multiple Contexts

CMA share many similarities with DSMT, especially dynamic thread generation and a ring topology used to communicate register values among contexts. However, the main differences between the two architectures are the following: (1) CMA requires special hardware mechanisms such as the multi-value cache, (2) CMA does not exploit nested loops, and (3) CMA is scalable but not based on a SMT core. Moreover, performance results of CMA, as well as some other previous architectures derived from the same research, were obtained with a trace simulator called ATOM [23, 9, 10, 11]. These results showed the potential performance benefit that can be obtained by exploiting only loops. However, since multi-threading exhibits dynamic behavior, trace simulators will not accurately reproduce misspeculations [2]. Moreover, as it will be shown later in this paper, frequent misspeculations cause significant degradation in performance. Our study of DSMT is based on a cycle-accurate, execution-based simulator capable of accurately reproducing the effect of misspeculations on the processor's performance.

Work closest to ours is Dynamic Multithreaded Architecture (DMT) [1]. DMT is designed around a SMT processor core. DMT also generates threads dynamically at run time and is capable of executing in parallel loops, procedures, and the code after the procedure. To relax the limitations imposed by register and memory dependencies, thread-level dataflow and data value prediction is used. A spawned thread uses as its input the register context from the thread that spawned it. Data speculation on the inputs to a thread allows new speculative threads to immediately start execution. Control logic keeps a list of the thread order and the starting PC of each thread. A thread stops fetching instructions when it reaches the start of the next thread in the order list. If for some reason a thread never reaches this point, it is considered misspeculated and consequently squashed. Threads communicate through registers and memory. Communication between threads is one way only and dictated by their order. Loads are issued to memory speculatively assuming that there are no dependencies with stores from previous threads. However, since threads do not wait for their inputs to be ready data misspeculation is common. DMT uses selective recovery on misspeculated instructions, which is initiated as soon as the correct input is available. Trace buffers outside the main pipeline hold all speculative instructions and their results. During recovery, instructions are fetched from the trace buffers and re-dispatched into the execution pipeline.

The major differences between DMT and DSMT are (1) DMT exploits procedures and loop continuations, (2) DMT employs multiple levels of speculation, (3) DMT employs more complex mechanisms for recovering from misspeculations. DMT and DSMT are complementary approaches to exploiting TLP. In particular, given that DMT exploits procedures and loop continuation code, integer applications will benefit more from the DMT architecture. In contrast, numerical applications consisting mainly of loops will execute more efficiently on DSMT.

## 3. DSMT Microarchitecture

Figure 1 shows the organization of the DSMT microarchitecture. Its core consists of a generic superscalar proces-sor organized into six pipelined stages: Fetch, Decode/Dispatch, Issue, Execute, Write-back, and Commit stages. The Fetch stage fetches a block of instructions from a thread in the usual manner, but can also fetch instructions from different threads based on the scheduling policy offered by the *Scheduler*. To support simultaneous execution of multiple threads, each thread has its own set of *Instruction Queue* (IQ), *Reorder Buffer* (ROB), and *Context*. Each Context represents the state of a thread and the multiple Contexts are also interfaced to the *Thread Creation and Initiation Unit* (TCIU), which controls how threads are cloned and executed. It also contains the *Loop Detection Unit*, which is responsible for detecting loops and supplying target addresses so that multiple threads can be cloned by TCIU. The following subsections highlight the functionality of the various components.

## 3.1 Loop Detection Unit

During the execution of a program, DSMT operates in either *DSMT* or *non-DSMT mode*. In the non-DSMT mode, there is only a single thread of execution and thus the processor behaves as a superscalar processor. When a loop is detected, the processor enters the *pre-DSMT* mode, and later if it is determined that multithreaded execution will improve performance it enters the *full-DSMT* mode. During pre-DSMT mode, the processor detects live registers and the information required to speculatively predict register values. In full-DSMT mode, the overlapped execution of loop iterations occurs. Moreover, there is always a single non-speculative context, which is the only thread permitted to clone speculative threads. This policy guarantees precise interrupts and reduces the complexity that would be required to control multiple speculative stages.

Whenever a taken backward-branch instruction is detected, branch and target addresses are recorded in the Loop Detection Unit. Later, if another branch instruction with the same branch target address is found, the processor enters the pre-DSMT mode. The structure of the Loop Detection Unit consists of a specially modified BTB augmented with additional fields to facilitate loop identification. In addition to the typical fields found in modern superscalar processors' BTB (e.g., branch address, target address, and branch prediction information), it contains the following information: (1) A flag indicating that the target address of this branch is the starting address of a loop; (2) the number of iterations that this loop has executed in the past (i.e., the number of consecutive taken branches); and (3) a type information indicating whether this is a *"good"* or *"bad"* loop for speculative execution based on its previous behavior.

Loop Detection Unit also contains a field that provides feedback on how loops behaved in their previous pre- and full-DSMT modes of execution. Four criteria are used to determine whether a loop is "good" or "bad" for speculative execution: (1) The number of iterations a loop executes, (2) the number of contexts currently available, (3) how much overlapped execution cloned threads exhibit, and (4) thread run-length. The first two criteria determine the potential TLP in the cloned threads. However, even in loops with a large number of iterations, it is possible that they may exhibit very low ILP during execution. This could be caused by the presence of a large number of inter-thread dependencies in the loop, or by frequent miss-speculation of loop iterations. In this case, the third criterion is used to determine the effectiveness of the DSMT execution mode. This criterion associates an Instructions-per-Clock (IPC) measure during the execution of a loop in DSMT mode. On the other hand, the fourth criterion together with the first two indicates how sustainable the overlapped execution can be. The first three criteria are combined to form the *sustained IPC* (SIPC) measure for a loop. A loop is labeled as good or bad based on a "break even" policy, where the observed IPC during DSMT execution is compared against the observed IPC for the non-DSMT execution of the same loop measured during pre-DSMT mode. If the IPC measured breaks even then the loop is labeled as a "good" loop for speculative execution. This way, we can guarantee that DSMT execution mode will result in as good or better performance than the non-DSMT mode of execution.

Nested loops provide opportunities for DSMT to select the appropriate thread granularity. For these loops, the SIPC measure is also used to select a particular loop in the nested loop structure that provides the best performance. The control of nested loop execution is handled by a special stack structure associated with Loop Detection Unit. This mechanism stores the branch and the target addresses of a loop in a stack of loops. Inner loops are stored at the bottom of the stack and outer loops at the top of stack. When a new loop is detected, its branch and target addresses are compared with the corresponding addresses stored at the top of the stack. If the new loop's branch and target addresses are in the range of addresses stored at the top of the stack, the loop is pushed onto the stack. Later during DSMT mode, the stack of loops is accessed based on the SIPC value obtained during the execution of each nesting level. The loop with highest SIPC in the nest is chosen, and all others are discarded.

## 3.2 Thread Creation and Initiation Unit and Multiple Contexts

The structure of TCIU and the multiple contexts are show in Figure 2. When the Loop Detection Unit detects a loop, using the policies described earlier, it latches the target address of the thread to be cloned to the Continuation register and sets the M-bit to indicate the processor is in pre-DSMT mode. TCIU also has a set of anchor bits called D_Anchor and R_Anchor bits, which are updated at the start of the second iteration with the status of D and R bits (which will be explained shortly), respectively, of the non-speculative context just before threads are cloned. These anchor bits provide a means of (1) speculating whether cloned threads should read the registers of their own context or from the registers of other contexts and (2) generating a new set of bits for future speculation. Afterwards, TCIU sends a signal to all other units indicating that full-DSMT mode of execution has started.

Table 2: DSMT Functional Unit Configuration

| FU Type | Int ALU | Int Mul | Int Div | FP Add | FP Div | FP Mul | Load/Store |
|---|---|---|---|---|---|---|---|
| # of FU | 8 | 2 | 2 | 2 | 2 | 2 | 2 |
| RSs | 8 | 2 | 4 | 4 | 2 | 4 | 4 |

Table 1: DSMT Configuration

| Inst. Queue size/context | L/S Queue size/context | ROB size/context | L1 Inst./Data Cache | L2 Unified Cache | Shared BTB size |
|---|---|---|---|---|---|
| 64 | 64 | 32 | 128KB/128KB 2-way | 256KB 2-way | 2KB 2-way |

In full-DSMT mode, the thread cloning process starts by copying the target address of a loop in the Continuation register to the PCs of each cloned context. To "jump start" each thread, immediate values of instructions of the form addi rd,rd,#immd are predicted and stored in the *Loop Stride Speculation Table* (LSST). The compiler generally uses rd as induction variables or to access data in regular patterns. The speculation used by DSMT is to predict the contents of a source register rd in LSST as rd=rd+iteration*immd, where iteration is the current iteration number. To avoid blind speculation on rd, each entry in LSST is associated with confidence bits based on 2-bit saturation counters. Next, the values stored in the non-speculative register file are copied to the *Registers* in each new context, and *Valid* (V) and *Speculative* (S) mode bits are set. The V-bit indicates that the context is valid and, therefore, the Fetch unit can start fetching instructions from its PC. The S-bit, when set, indicates the context is running speculatively. Therefore, a single non-speculative context owns the precise state of the processor in its private register file.

Each context is also equipped with a register file that provides with a distinct logical view of its state, allowing fast register access in a context. The multiple contexts are interconnected in a ring fashion, and the Head and Tail registers of TCIU determine the first and the last thread currently running on the DSMT processor.

To keep track of inter-thread register dependencies, each register is associated with a set of utility bits (in addition to the usual ROB entry tags and Busy bits found in the register file of modern superscalar processors). They are:

*Ready* (R) bit – When set, it indicates some instruction(s) logically preceding this one in the thread's program order has committed a value to the register; otherwise, no value has been committed to the register and there are no instruction(s) in the local ROB that will commit to this register. R bits reflect whether registers can be read from their own context or speculatively read from a predecessor context. This flag also indicates that successor contexts need look no further than this context to get the value they need.

*Dependency* (D) bits – Keeps track of registers that have inter-thread dependencies. When a register is read, if its R-bit is zero and there are no other instructions in the ROB that will commit to the register, a check is then made to see if its R_Anchor-bit is set. If both of these conditions are true (i.e., register was not written in the current context but other context previously wrote to that register), it means inter-thread dependence exists for the register and the D bit is set. These bits will serve as D_Anchor bits to facilitate speculation on how registers are accessed.

*Load* (L) bit – When set, it indicates that the register has been speculatively read from a predecessor context. When an instruction attempts to read a register in its own context, L-bit is set if its R-bit is zero and there is no instruction(s) in the ROB that will commit to this register. If later on, it is determined that a register was actually written in a context when the successor thread L-bit is set, then all successor threads are squashed. Since these register reads are speculative, a confidence based on 2-bit saturation counter is associated with each register and stored in the *Register Read Confidence Table*.

Each context has an associated *Join* (J) bit, which is set to indicate that an iteration has completed. This bit is used to synchronize multiple threads that have reached this state. At this point, speculative contexts must wait until the non-speculative context commits its results and transfers the non-speculative register values to a new context. Transferring the flag from the non-speculative context comprises copying the value of each register in the non-speculative context register file to the new non-speculative context. However, the copying process skips those registers values that were identified as live during the execution of the new non-speculative context.

At this point, there are two important implementation details in the DSMT architecture that need to mentioned. First, before threads can be cloned, a couple of loop iterations must be executed in non-DSMT mode to establish the contents of the Continuation register, LSST entries, and D_Anchor and R_Anchor bits. Second, all cloned threads execute speculatively. Thus, when the non-speculative thread completes, its successor thread becomes the new non-speculative thread. Therefore, D_Anchor and R_Anchor are updated with the values of the R and D bits of the just completed, non-speculative thread and the Head register is updated to point to the new non-speculative thread.

## 3.3 Resolution of Inter-thread Dependencies in Registers and Memory

Register dependencies between iterations are resolved by speculatively accessing registers based on the D_Anchor bits. If R-bit is set for a register, the register value can be read directly from its own context. Otherwise, first-level speculation, called *register dependence speculation*, is performed based on its D_Anchor bit to determine which thread the value should be read from. For example, when an instruction in a thread tries to read its own register with its R-bit equal to zero, the D_Anchor bit for the register is checked. If the D_Anchor bit is set, it indicates that previous executions of the iterations had an inter-thread dependency on the register. Therefore, the speculation assumes that this inter-thread dependency will likely exist in the current execution of speculative threads, and the register, when ready (i.e., R-bit = 1), is read from its immediate predecessor thread. If the register dependence speculation turns out to be wrong, due to dynamic behavior of loop iterations, and the immediate predecessor thread does not generate the register value, the second-level speculation is used. This involves searching back for the last thread that generated a value for the register.

On the other hand, if the D_Anchor bit of a register is zero, it indicates that previous executions of the loop iterations did not have an inter-thread dependence on the register. However, since dynamic behavior of loop iterations may have changed a register due to an inter-thread dependent register, the speculation used is to assume that predecessors may have modified the register and the register is read from the last thread that wrote to this register. If no predecessor threads have modified the register, it is read from its own context.

Since the DSMT processor relies very heavily on speculation, the cloning and speculative execution of threads require a method to detect and squash threads when misspeculation occurs. The detection of misspeculation is performed when registers are written in each context during the Commit stage. Whenever a thread writes to a register, the L bits of the successor threads are checked to see if any thread has read the register earlier. If so, that thread and all of its successor threads are squashed and reinitiated.

In order to ensure the proper ordering and yet maximize the overlapped execution of the cloned threads, a new iteration is initiated in the context whenever any other completes. Thus, when a thread completes a single iteration, that context will set the appropriate J-bit in the TCIU. Since the just completed iteration has properly updated its R and D bits, these bits become the new set of anchor bits. In addition, the just completed thread's successor now becomes the non-speculative thread. Therefore, TCIU can reinitiate the next iteration by appropriately changing the Head and Tail registers and cloning a new thread.

In DSMT, loads from different threads, either speculative or non-speculative, can be executed speculatively. However, only the non-speculative threads are allowed to perform stores. To ensure that the sequential semantics is not violated, the *Memory Dataflow Resolution Table* (MDRT) is used. Load/store operations are kept in one of the *Load/Store Queues* (LSQs) according to its tag. In addition to allowing loads to bypass stores and forwarding values from stores to loads that have been disambiguated, the LSQ also acts as a buffer so that stores can speculatively commit locally. This prevents uncommitted stores from speculative threads from blocking the ROB. A special logic selects load/store operations, giving priority to those corresponding to the non-speculative context, and forwards them to the memory subsystem. MDRT checks these operations to ensure the correct state of the memory. MDRT is a fully associative buffer with each entry containing a valid (V) bit, a word address *(addr)*, and a value. In addition, each thread has an L-bit and an S-bit indicating whether the memory word has been loaded or stored, respectively.

Loads can proceed normally for non-speculative threads. However, speculative threads performing a load, check to see if an entry exists based on *addr*. If none exists, an entry is allocated for the *addr*. If an entry is found, then its L-bit is set and the load is allowed to precede its execution. A store is not allowed to update the memory unless it is from a non-speculative thread and it is at the head of its ROB. This guarantees that the precise state of the processor can be maintained. When a non-speculative thread performs a store, it sets the S-bit for that thread. In addition, it checks to see if other threads have their L bits set. If any thread has read this memory value too early, the thread and all of its successor threads are squashed

## 4. Simulation Environment

To evaluate the performance of DSMT, DSMTSim was implemented. DSMTSim is a cycle-level accurate, execution-driven simulator capable of operating in different execution modes, which are: (a) fast, in-order simulation and (b) detailed wide-issue, out-of-order, multiple context simulation. The fast simulation mode allows quickly placing the simulator in a particular section of a benchmark code, skipping non-representative parts like those corresponding to initialization. During fast mode simulation, instructions are read directly from memory and executed in sequence. Conversely, in the detailed simulation mode, all the memory hierarchy and the pipeline stages of the simulator are exercised. The simulator loads a binary program into its internal memory, using the detailed simulation mode, and then simulates, cycle-by-cycle, all the processing performed by the main pipeline.

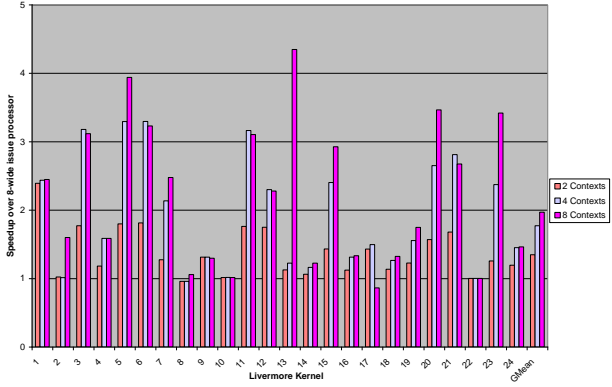

Figure 3: DSMT performance based on ICount2.8-modified fetch policy.

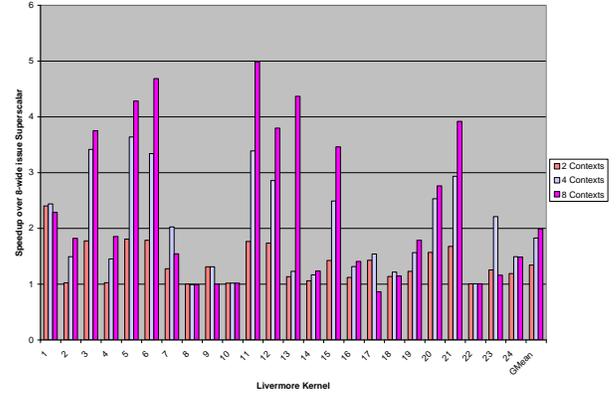

Figure 4: Performance with unlimited number of fetch ports.

DSMTSim executes PISA binaries generated from SimpleScalar's GCC. However, except for the memory model and the syscall support, DSMTsim is a completely new simulator that shares very little with the SimpleScalar's sim-outorder simulator. One of the main advantages of DSMTSim is that it reproduces in detail all the dynamic events that the DSMT's multithreaded execution mode generates. Specifically, DSMTSim processes instructions out-of-order, spawns new threads, synchronizes threads, and flushes and recovers from misspeculated threads. At intra-thread level, branches are predicted and executed speculatively. Also, register values are generated and sent from producer instructions to consumers either at intra-thread or inter-thread level *on-the-fly*. On-the-fly value passing closely resembles the action performed by real superscalar processors, and is also used as a means for checking the correct operation of the tagging and out-of-order execution mechanisms included in the simulator. In this way, correct manipulation of data values is ensured during speculative execution.

## 5. Performance Simulation Results

Our simulation study of DSMT was based on Livermore loops and SPEC95 benchmarks. Livermore loops consist of 24 core calculations used in familiar numerical algorithms, such as matrix multiplication, Cholesky's conjugate gradient, Monte Carlo's search, etc. Both sets of benchmarks were compiled without modifications to the original C source code, using GCC with the -O3 optimization flag turned on. In this simulation, the I-cache is two-ported [27] and the fetching policy used by the Scheduler (see Figure 1) is similar to ICount2.8 [24]. ICount2.8 policy employs two fetch ports, each one able to fetch up to eight instructions per clock cycle. Using this policy, priority is given to the context with the lowest number of instructions (ICount) in its decode, rename, and issue stages. However, SMT's original ICount2.8 policy was slightly modified for DSMT (called *ICount2.8-modified*) by selecting first the non-speculative thread, and then choosing among the rest of speculative threads the thread with the lowest ICount. DSMTSim's simulation parameters and the functional unit configuration were based on Table 1 and 2. Also, each context was allowed to issue up to four instructions per clock cycle.

We first ran Livermore loops to assess the effectiveness of DSMT on benchmarks that contains mainly loops. Figure 3 shows the speedup obtained by the DSMT processor with 2, 4, and 8 contexts when compared to a single context 8-wide issue superscalar processor. In the graph, *GMean* represents the geometric mean of the speedup obtained by all 24 kernels. As the figure shows, the maximum speedup obtained by DSMT was on average 34%, 84%, and 100% with 2, 4, and 8 contexts, respectively. Figure 4 shows the performance of an ideal case where the number of fetch ports is equal to the number of contexts. The difference in performance between the two configurations (ideal and ICount2.8-modified) is on the average 0%, 5%, and 3% for 2, 4, and 8 contexts, respectively. Therefore, the performance obtained by using two fetch ports with ICount2.8-modified policy is very close to the one obtained by the ideal configuration.

These results also show that the speedup was negligible for certain loops. Statistics gathered from DSMTSim showed that the LSST mechanism resulted in a high thread misprediction rate for Kernel-8 (integration). This is because the memory access pattern in this loop is complex, and therefore, the simple value prediction method for LSST is unable to accurately predict the induction variables used to control the execution of the loop. In contrast, low performance obtained in Kernel-10 (difference predictor) is due to a combination of two factors: LSST's low thread prediction rate and high thread synchronization rate. In DSMT, a speculative thread is blocked when it finishes its execution before the non-speculative context does. Since the non-speculative context is the only one capable of enabling other speculative contexts to become non-speculative, the gap produced when the non-speculative thread is delayed causes degradation in performance. The factor that caused performance loss in Kernel 22 (Plankian distribution) is due high branch misprediction rate.

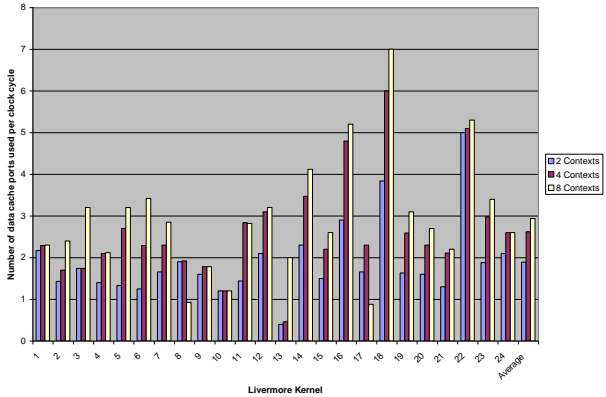

Figure 5: Data cache accesses per clock cycle

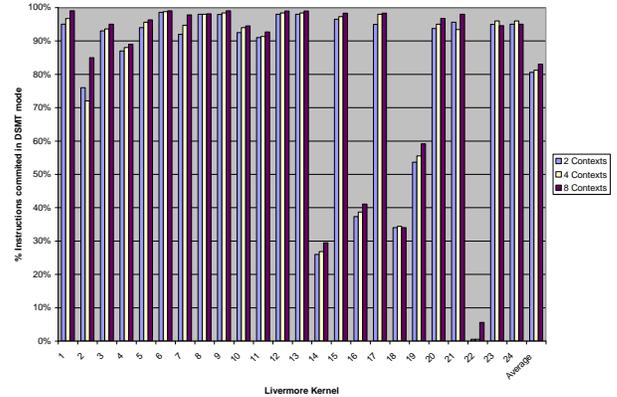

Figure 6: Instructions committed in DSMT mode for Livermore loops

Figure 3 also indicates that DSMT does not always obtain the best performance using the maximum number of threads. The reason for this is two-fold. First, since each context executes nearly the same code and thus resource requirements of each thread are very similar during each iteration. Therefore, the likelihood of resources conflicts increase as more contexts are spawned. This is especially critical in loops with a very large number of numerical calculations and/or a large number of memory accesses, such as Kernel 8 (integration), and Kernel 10 (difference predictor).

Dynamic behavior within loops is the other main cause for the poor performance exhibited by some loops with eight contexts. Loops with dynamic behavior tend to generate higher inter-thread misspeculation. Misspeculation occur mainly when register values are read too early by the speculative threads, as is the case of Kernel-17 (conditional computation), which contains five *goto* statements in the loop body. In this loop, misspeculated threads that are squashed cause a decrease in performance for eight contexts. An analysis of the code show that some internal backward jumps are mistaken as loops by the Loop Detection Unit. However, whether these backward branches are taken or not depends on the internal variables in the loop. Therefore, using the maximum number of contexts increases the probability that many threads will be squashed when the conditional branch is frequently not taken.

Figure 5 shows the average number of data cache ports accessed per clock cycle with Livermore loops. As expected, with more contexts in the processor more pressure is exercised on the data cache ports every cycle. However, Kernel-8 and Kernel-17 reveal a different pattern in accessing the data cache: With 8 contexts less data cache ports are used. The reason is LSST misspeculations occur very often (especially in Kernel-17) due to dynamic behavior in the loop. Therefore, with more contexts, more threads are squashed before the memory operations could reach the load/store ports, which reduces the pressure on the data cache but degrades performance. These results suggest that the number of data cache ports required by DSMT is around three. However, to take into account the more demanding memory requirements of more complex benchmarks, DSMT uses four ports. It is interesting to note that reducing the fetch bandwidth (using ICount2.8-modified) also reduces the pressure on the data cache ports. This indicates that there is a tradeoff between increasing the fetch bandwidth to improve performance, but also limiting it if misspeculations occur very often.

Figure 6 shows the percentage of instructions committed in DSMT mode for each of the Livermore loops. On the average nearly 80% of the instructions were committed in DSMT mode, which indicates the Loop Detection Unit is very effective in identifying and exploiting these loops. As Figure 6 shows, DSMT is able to find a significant amount of TLP in Kernel-8 and -10. However, as Figure 3 shows the speedup obtained in these loops is negligible due to thread misspeculation. In contrast, DSMT is unable to find enough TLP in Kernel-22 due to a very high branch misprediction rate. Statistics from DSMTsim also show that Kernel-14, -16 and –18, which have several internal loops and *if-then-else* statements in the loop body, produce a relatively high number of branch mispredictions.

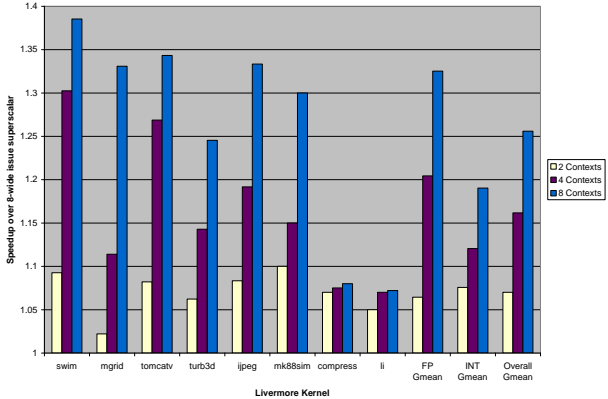

Figure 7: DSMT performance using SPEC95 benchmarks

The Livermore loops were used to analyze and explore part of the design space of DSMT and to characterize its dynamic behavior. However, the real advantage of DSMT is observed when more complex applications are executed. Figure 7 illustrates the performance of DSMT running several SPEC95 benchmarks. In the simulations, 500 million of instructions were executed, but the first 200 million instructions corresponding to code initialization were skipped using the fast simulation mode provided by DSMTSim. The reference inputs of the SPEC95 benchmarks were used during all simulations. Figure 7 shows the performance results obtained by DSMT for both the SPEC95-FP and SPEC-Int benchmarks. Results show the average speedup obtained by DSMT for all benchmarks is 7%, 16.5%, and 26% for 2, 4, and 8 contexts, respectively.

As other previous studies have found [9, 10, 23], in the DSMT architecture SPEC95-FP benchmarks provide the better speedup (32.5% on average with 8 contexts). The reason is that essentially they contain many more loops than SPEC95-Int. SPEC95-Int obtained an average speedup of 19% average speedup with 8 contexts. As these results show, numerical applications will benefit more from the DSMT model than non-scientific applications.

## 6. Conclusions and Future Work

This paper presented the DSMT architecture and its performance results. DSMT employs aggressive forms of speculation to dynamically extract TLP and ILP from sequential programs. Unlike other similar architectures, DSMT uses simple mechanisms to synchronize threads and keep track of inter-thread dependencies, both in registers and memory. The novel features in DSMT include (1) using information obtained during the sequential execution of code segments as a hint to speculate the subsequent behavior of multiple threads, and (2) utilizing a greedy approach that chooses sections of code that are more likely to provide the highest performance based on its past dynamic behavior.

The performance results of DSMT were obtained using a cycle-accurate, execution-based simulator DSMTsim, which is capable of executing mispredicted paths of execution, jointly with run-time generation, control and synchronization of multiple threads. Our simulation results show that speculative dynamic multithreading based on extracting threads from loops has very good potential for improving SMT performance, even when only a single program is available for execution. DSMT obtained on average nearly 100% speedup executing the Livermore loops and 26% of improvement for SPEC95 benchmarks. However, the performance improvement obtained by DSMT is limited for non-numerical applications when only loops are exploited. The reasons for this are two-fold. First, some applications simply lack TLP as well as ILP within a thread. This limits the performance achievable by DSMT and other similar architectures. Second, as our simulations results have shown, the dynamic behavior of speculative multithreading causes frequent mispredictions in some loops that may produce a detrimental effect on DSMT's performance. Therefore, the challenge is to maximize the amount of TLP that DSMT is capable of exploiting and at the same time reduce the frequency of misspeculations.

There are a number of ways the DMST architecture can be improved. First, as [13] suggests, exploiting both procedures and loops will likely be required to improve performance. Second, a more sophisticated value prediction mechanism will be needed. Another important bottleneck observed during our study was the memory dataflow mechanism. Long running threads that access a large number of memory locations may cause MDRT to quickly fill-up when there are insufficient number of data cache ports and thus cause the entire pipeline to backup. This was the reason why the Loop Detection Unit did not choose the outer-loop in some of the benchmarks, such as Kernel-21 (matrix multiplication), to clone threads. However, even when the middle-loop is chosen, its speculative threads may generate a large number of loads and stores, especially stores that cannot commit. This will cause the MDRT to backup and the result of this bottleneck percolated all the way back up to the IQs. Therefore, a method that "throttles" the thread execution is needed to avoid filling up the MDRT too quickly.

In addition, choosing the optimum number of contexts for execution may be critical for some applications. However, since selecting the optimal number of contexts at run-time is difficult, one possible solution is to carry out this information from the compiler to the architecture using a technique similar to the one described in [14]. Finally, improving the branch prediction mechanism employed by DSMT, which is currently based on a simple 2-bit saturating counter, will also be crucial.